\documentclass[journal]{vgtc}         

\ifpdf
  \pdfoutput=1\relax                   
  \usepackage{graphicx}                
  \DeclareGraphicsExtensions{.pdf,.png,.jpg,.jpeg} 
\else
  \usepackage{graphicx}                
  \DeclareGraphicsExtensions{.eps}     
\fi%

\graphicspath{{figures/}{pictures/}{images/}{./}} 

\usepackage{microtype}                 
\PassOptionsToPackage{warn}{textcomp}  
\usepackage{textcomp}                  
\usepackage{mathptmx}                  
\usepackage{times}                     
\usepackage{cite}                      
\usepackage{tabu}                      
\usepackage{booktabs}                  
\usepackage{enumerate}
\usepackage[shortlabels]{enumitem}
\usepackage{caption}
\usepackage{subcaption}

\onlineid{0}

\vgtccategory{Research}

\vgtcpapertype{please specify}

\title{Design and Evaluation of Scalable Representations of Communication in Gantt Charts for Large-scale Execution Traces}

\author{Connor Scully-Allison, Katherine E. Isaacs}
\authorfooter{
\item
 Connor Scully-Allison, University of Arizona. E-mail: cscullyallison@email.arizona.edu.
\item
 Katherine E. Isaacs, University of Arizona. E-mail: kisaacs@cs.arizona.edu.
}

\abstract{
    Gantt charts are frequently used to explore execution traces of large-scale parallel programs found in high-performance computing (HPC). In these visualizations, each parallel processor is assigned a row showing the computation state of a processor at a particular time. Lines are drawn between rows to show communication between these processors. When drawn to align equivalent calls across rows, structures can emerge reflecting communication patterns employed by the executing code. However, though these structures have the same definition at any scale, they are obscured by the density of rendered lines when displaying more than a few hundred processors. A more scalable metaphor is necessary to aid HPC experts in understanding communication in large-scale traces. To address this issue, we first conduct an exploratory study to identify what visual features are critical for determining similarity between structures shown at different scales. Based on these findings, we design a set of glyphs for displaying these structures in dense charts. We then conduct a pre-registered user study evaluating how well people interpret communication using our new representation versus their base depictions in large-scale Gantt charts. Through our evaluation, we find that our representation enables users to more accurately identify communication patterns compared to full renderings of dense charts. We discuss the results of our evaluation and findings regarding the design of metaphors for extensible structures.
} 

\keywords{}

\CCScatlist{ 
 \CCScat{K.6.1}{Management of Computing and Information Systems}%
{Project and People Management}{Life Cycle};
 \CCScat{K.7.m}{The Computing Profession}{Miscellaneous}{Ethics}
}

\teaser{
  \centering
  \includegraphics[width=\linewidth]{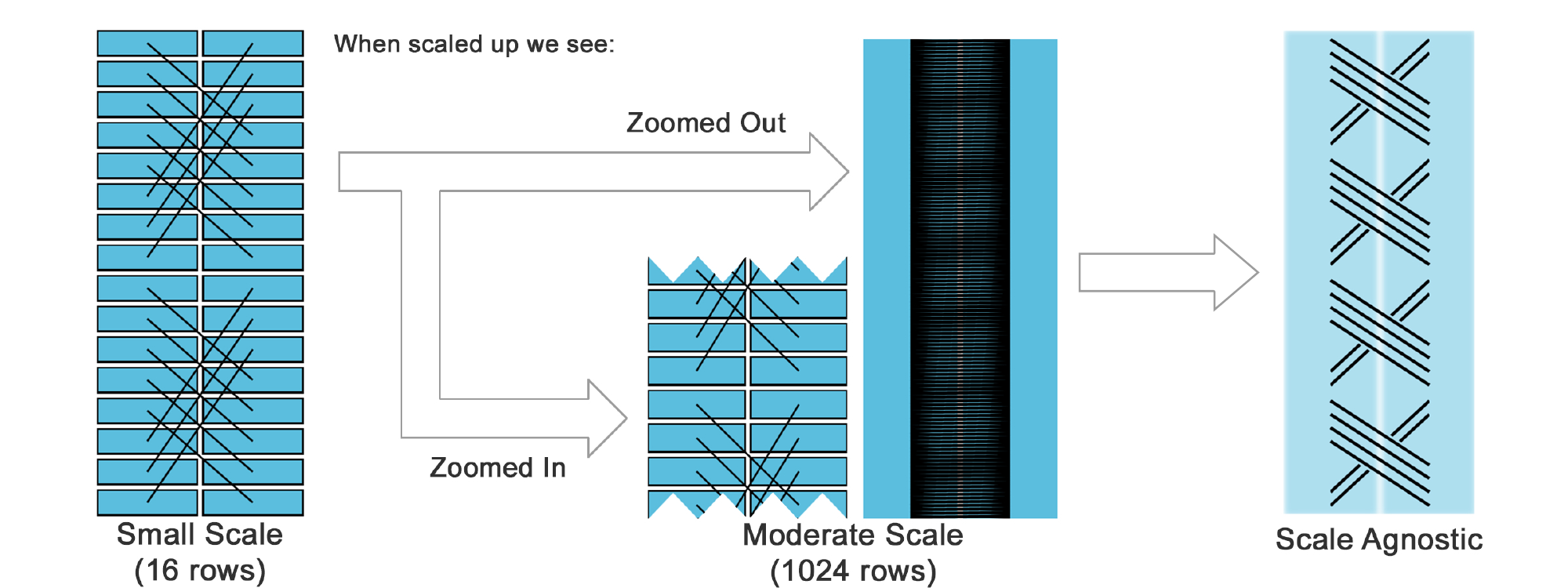}
  \caption{Communication between rows in a Gantt chart are interpretable at small scales but not further, despite being highly regular. We study how people identify and distinguish similar structures at different scales and design a scale-agnostic representation.}
  \label{fig:teaser}
}

\vgtcinsertpkg

\begin{document}

\firstsection{Introduction}

\maketitle

Large-scale parallel computation is required to achieve meaningful results from scientific simulations that model domains like climate change, medicine, and energy. High performance computing (HPC) is concerned with making such computation possible. A key strategy is performance analysis and optimization. 

\begin{figure}[htb]
    \centering
    \includegraphics[width=\columnwidth]{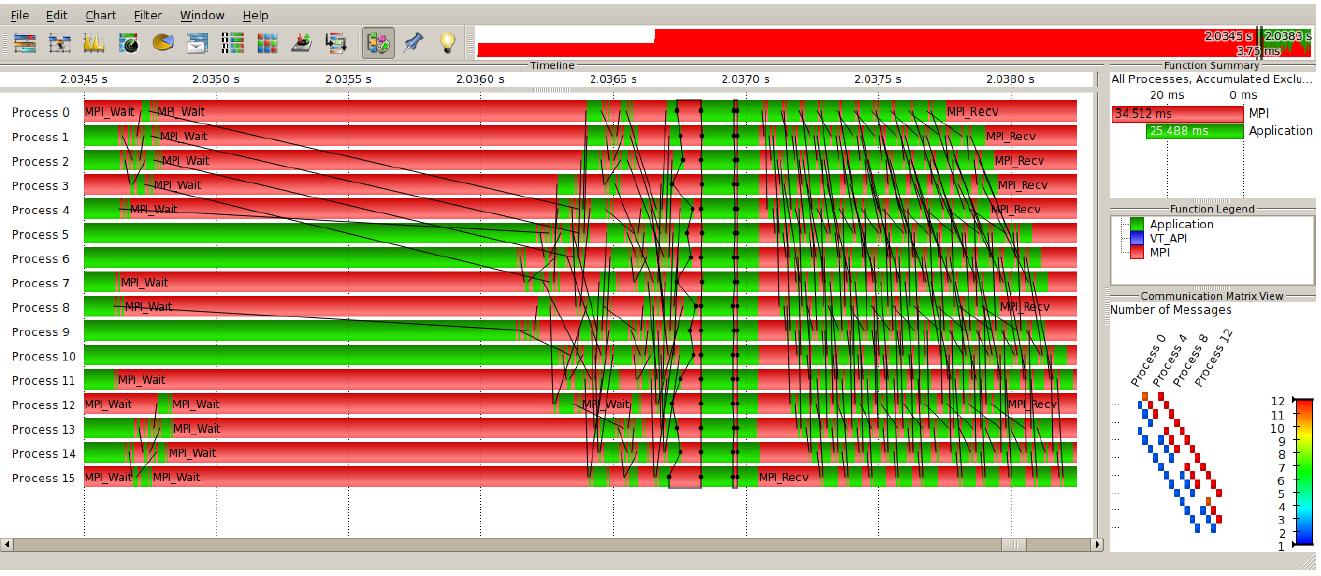}
    \caption{A Gantt chart produced by Vampir from \cite{isaacs2014state}, showing a slice of time in a 16-PE execution. Black lines denote communication.}
    \label{fig:simple_gantt}
\end{figure}

One approach in performance analysis is collecting and analyzing an execution trace---a record of events occurring during program execution. Such traces are commonly visualized with Gantt charts (\autoref{fig:simple_gantt}). Each processing element (PE), e.g., a hardware thread, is assigned a row. Function calls are drawn as rectangles over their time (x-axis) interval. 

PEs must communicate to exchange data and results. This communication is represented by overlaid lines in the Gantt chart, diagonally sweeping the PE space. Even in tiny parallel programs, i.e. those executed on 32 PEs, the depiction can become cluttered. Arranging events on a {\em logical time} axis, where events are shown based on logical relationships, rather than a physical time one, has been shown to aid in understanding the data~\cite{isaacs2014combing}. Events that ideally would have occurred at the same time are easier to compare and patterns in the communication structure are revealed.

\begin{figure}
    \centering
    \begin{subfigure}{0.15\textwidth}
        \centering
        \includegraphics[width=\textwidth]{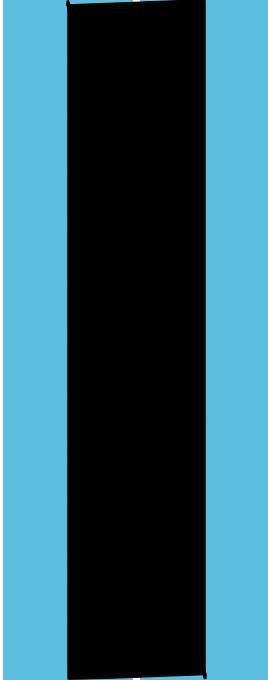}
        \caption{Continuous ring}
        \label{fig:dense-ring}
    \end{subfigure}
    \begin{subfigure}{0.15\textwidth}
        \centering
        \includegraphics[width=\textwidth]{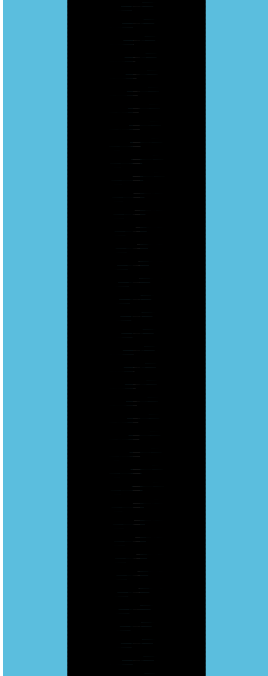}
        \caption{Continuous offset}
        \label{fig:dense-offset}
    \end{subfigure}
    \begin{subfigure}{0.15\textwidth}
        \centering
        \includegraphics[width=\textwidth]{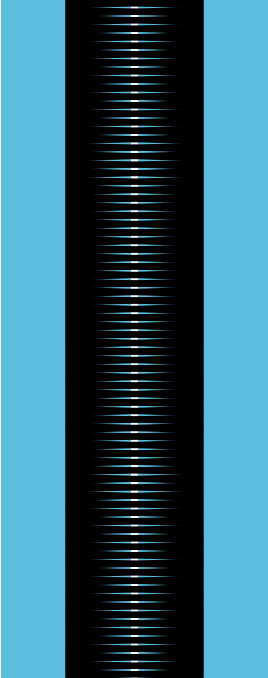}
        \caption{Exchange}
        \label{fig:dense-exchange}
    \end{subfigure}

    \caption{Examples of dense communication lines in Gantt charts across two time steps. The resolution is such that each processing elements is less thann a pixel and individual lines occlude. }
    \label{fig:dense_gantt}
\end{figure}

However, logical time encodings retain the vertical scaling problem of Gantt charts. A modestly sized program executed on 1,024 PEs may induce so much occlusion that communication lines appear as a solid black shape. \autoref{fig:dense_gantt}, shows three examples with different communication structures that are difficult to identify and distinguish. These are all manifestations of code designed to cover the number of PEs allocated and thus often exhibit repetitive structure. We seek to understand how potential performance analysts interpret these structures in Gantt charts and what factors they use to differentiate them so scalable representations can be designed.

We conduct a qualitative study at an HPC conference to investigate salient features in existing depictions and how people discern patterns, especially in cases where the source code is the same, but the instance differs due to PE count. Leveraging our findings, we decompose these patterns into domain-specific traits and design encodings for communication pattern representations. We then conduct a controlled user study of these designs versus the standard depictions. The results show the new designs enable users to more accurately identify key pattern traits and suggest further directions in Gantt chart research. Finally, we reflect upon the nature of the concept of patterns and designing them, recruiting participants in high-expertise domains, and speculative visualization design. 

In summary, our contributions are:

\vspace{-0.5ex}

\begin{itemize}
    \itemsep0em
    \item the design (\autoref{sec:design}) of scalable representations of communication patterns for Gantt charts, 
    
    \item evaluations (\autoref{sec:prelim}, \autoref{sec:analysis}) of efficacy of dependency depiction in Gantt charts, and
    
    \item reflections (\autoref{sec:reflections}) on visualizing patterns, participant recruitment, and speculative visualization design. 
    
\end{itemize}

We first present relevant background (\autoref{sec:background}) and related work (\autoref{sec:related}). We conclude in \autoref{sec:conclusion}.


\section{Background}
\label{sec:background}

We discuss relevant background in high performance computing, execution traces and their visualization, and communication patterns.

\begin{figure*}
    \centering
    \begin{subfigure}{0.18\textwidth}
         \centering
         \includegraphics[width=\textwidth]{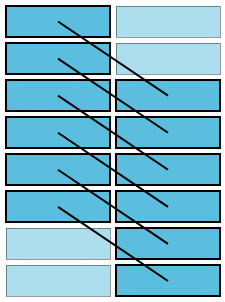}
         \caption{Continuous offset pattern}
         \label{fig:oc}
    \end{subfigure}
    \begin{subfigure}{0.18\textwidth}
         \centering
         \includegraphics[width=\textwidth]{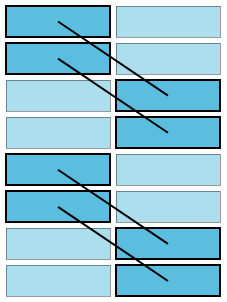}
         \caption{Grouped offset pattern}
         \label{fig:og}
    \end{subfigure}
    \begin{subfigure}{0.18\textwidth}
         \centering
         \includegraphics[width=\textwidth]{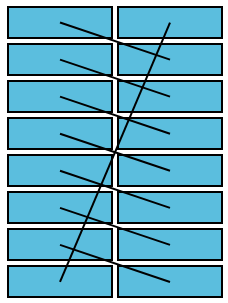}
         \caption{Continuous ring pattern}
         \label{fig:rc}
    \end{subfigure}
    \begin{subfigure}{0.18\textwidth}
         \centering
         \includegraphics[width=\textwidth]{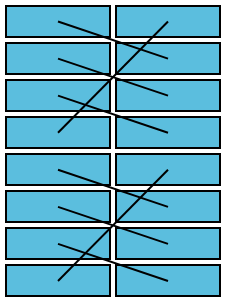}
         \caption{Grouped ring pattern}
         \label{fig:rg}
    \end{subfigure}
    \begin{subfigure}{0.18\textwidth}
         \centering
         \includegraphics[width=\textwidth]{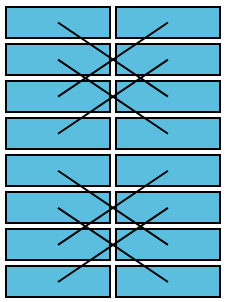}
         \caption{Exchange pattern}
         \label{fig:eg}
    \end{subfigure}
    \caption{Communication patterns supported by our new design. Each of these show 8 PEs and two time steps.}
    \label{fig:commpatterns}
\end{figure*}

\vspace{1ex}

\textbf{High Performance Computing (HPC).} A central goal of HPC is completing large computational problems, such as scientific simulations or large numeric (e.g., matrix) calculations, more efficiently, typically through the use of large-scale parallel resources. In some cases, the computational problem is large enough that it is infeasible to compute on fewer resources. To perform such computation, the problem is typically divided into smaller sub-problems, such as slices of the domain, and distributed across the parallel resources. We refer to these parallel resources as {\em processing elements (PEs).}

During program execution, there are times when the PEs must pass data, such as partial results or shared boundary information, to each other. We refer to the messages sent between PEs as {\em communication.}

\vspace{1ex}

\textbf{Execution Traces.} As HPC aims for efficient use of resources, analyzing a program for its performance (e.g., the time it took to execute) is a common task. At its most simple, performance measurement might just be recording the time taken for a full execution. While this tells developers and performance analysts whether a version of the code or a change in configuration performed better than others, it doesn't reveal why. {\em Tracing} is a form of performance measurement that logs individual timestamped events, such as the beginning and ending of functions or communication calls. From these fine-grained measurements, the history of a program's execution can be reconstructed and used to explore how events, in concert, led to observed performance.

\vspace{1ex}

\textbf{Visualizing Execution Traces with Gantt Charts.} Traces are often employed in an attempt to understand complex behavior from the intersection of the parallel program and the resources on which it was executed. This exploratory task is often performed with the aid of a Gantt chart. In HPC, typically the x-axis denotes time and processing elements are stacked as rows in the y-axis, usually in order of their ID. Rectangles are plotted to show the start, end, and executing PE of a function call. Straight lines originating at one rectangle and spanning rows to another show the send time, receive time, and end points of communication. \autoref{fig:simple_gantt} shows a time-slice of a small-scale (16 PEs) execution trace, rendered by the commercial software Vampir~\cite{nagel1996vampir}.

\begin{figure}
    \centering
    \includegraphics[width=\columnwidth]{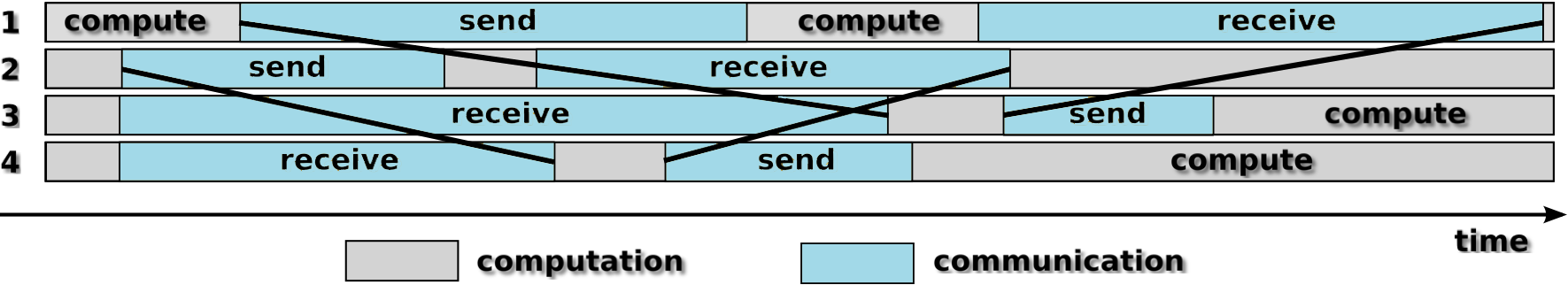}\\
    \vspace{1ex}
    \includegraphics[width=\columnwidth]{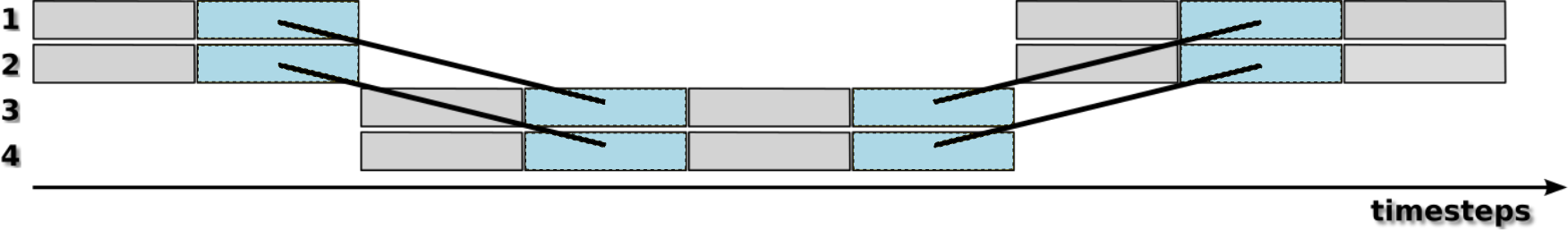}
    \caption{A toy example trace visualized in a Gantt chart in physical time (top) and idealized unit time (bottom). The latter enforces logical rules, like receives happening after sends, and aligns events on PEs created by the same code.}
    \label{fig:physical-logical}
\end{figure}

While most Gantt charts in HPC depict physical time in seconds (\autoref{fig:physical-logical} (top)), another approach is to arrange events into an idealized unit time using logical rules~\cite{Leblanc1990, Schaubschlager2003DeWiz, Isaacs2015, Isaacs2016} (\autoref{fig:physical-logical} (bottom)). In this scheme, events are partitioned by relationships, e.g., functions called by a PE remain in that order and sends must occur before receives~\cite{Lamport1978}. Physical time data is then encoded with another channel, such as color. This arrangement has been used to emphasize what programmers might expect in an ideal situation, where work is evenly divided and there is no contention for resources, thereby allowing them to identify code regions and compare events that {\em should have} behaved similarly. Use of these views has resulted in significant performance improvements in real-world production code~\cite{isaacs2014combing}.

Typically the main view of a Gantt chart shows a navigable window of time. The bounds of the window may also be adjusted by brushing an overview (\autoref{fig:simple_gantt} top left). As the number of PEs increases, vertical panning and zooming can be used. When zoomed in, only some of the communication lines are shown. When zoomed out, the lines are either omitted or obscure the view. The goal of our work is retaining recognizability of communication at larger scales.

\vspace{1ex}

\textbf{Communication Patterns.} In many parallel programs, processing elements communicate at logically equivalent times with neighbors that can be calculated based on their ID, the total number of PEs, and optionally additional information about the domain. For example, at a specific line of code, all PEs might send a message to the PE that is their own ID plus one. We refer to commonly observed instances of these as {\em communication patterns.}

Communication patterns often show a degree of regularity and symmetry when depicted in idealized unit time. When recognized in a trace visualization, communication patterns can serve as an indicator of what part of the source code is being executed and what the program is trying to accomplish. Thus, we focus this work on preserving that capability. We describe the specific patterns we consider (shown in \autoref{fig:commpatterns}):

\begin{enumerate}
    \itemsep0em
    \item \textbf{Offsets.} Each PE sends to the PE that is their own ID plus some offset or {\em stride} if (ID + stride) is a valid PE ID. These appear as parallel lines in idealized time Gantt charts. These are seen in applications like PF3D~\cite{pf3d} and several NAS benchmarks~\cite{nas}.
    
    \item \textbf{Rings.} Each PE sends to the PE that is their own ID plus some stride, modulo the total number of PEs in the ring. In idealized time Gantt charts, these appear like offsets with cross lines where the send-receive pair ``wraps around''. These are used in Jacobi iterations, libNBC~\cite{libnbc}, and BT from the NAS benchmarks~\cite{nas}.
    
    \item \textbf{Exchanges.} The PEs are paired off, each one sends to their partner. Typically the pairs are regularly spaced. These are identical to a ring where the stride is half the ring size. We separate this class however as they are thought of separately and are typically formed of many small groups across the PE ID space. These seen in LULESH~\cite{LULESH} and MG from the NAS benchmarks~\cite{nas}.
    
    \item \textbf{Stencils.} The PEs are assumed to be arranged in a kD (typically 3D) grid. Each PE sends to all of its j-hop neighbors in the grid, sometimes including diagonal neighbors. Stencils are used in applications such as AMG~\cite{amg} and LASSEN~\cite{lassen}.
    
\end{enumerate}

Offsets and rings can subsume the entire ID space. Alternatively the ID space can be partitioned and the calculation of the receiving ID can be done within that partition. These form {\em groups} in the pattern.

We note that communication patterns are {\em extensible}, much like the programs that contain them. Executing using more PEs can results in a larger instance or more replications of the same pattern. In a sense, a pattern defines a family of structures that vary across number of PEs, similar to how there are families of graphs, such as cliques. We use this fact to design visualizations of these patterns and test their efficacy.

\section{Related Work}
\label{sec:related}

We discuss relevant related work in visualizing execution traces and extensible structures.

\vspace{1ex}

\textbf{Visualizing Execution Traces with Many Processing Elements.} Most approaches for scaling Gantt charts for execution traces do so by removing communication entirely. Zinsight~\cite{Zinsight} elides PEs on the same chip while Cottam et al.~\cite{Cottam2015} use pixels in each row to show event-type distribution. Muelder et al.~\cite{Muelder2009} and Sigovan et al.~\cite{Sigovan2013} forgo separate rows for PEs, transforming the y-axis into duration and plotting all PE events in the same space. Ocelotl~\cite{Dosimont2014Ocelotlb} aggregates events that are similar in time and PE, a strategy similar to LiveGantt~\cite{Jo2014} which was used for manufacturing schedules rather than HPC.

SmartTraces~\cite{Osmari2014SmartTraces} combines line-less Gantt charts with a node-link diagram of tasks to show overall workflow, but not communication. Fine-grained task graphs have been used in lieu of Gantt charts, but aggregation schemes~\cite{Huynh2015DAGViz, Reissmann2017GrainGraphs} require fork-join programming models. 

Few approaches retain communication lines in Gantt charts. Ravel~\cite{isaacs2014combing} shows subgraphs induced by communication lines, but it is unclear if these are large enough to show patterns. Brendel et al.~\cite{Brendel2016} apply force-directed edge bundling, but only for small PEs counts.

\vspace{1ex}

\textbf{Visualizing Extensible Structures.} Biological workflows often have repetitive subgraphs. AVOCADO~\cite{Stitz2016} collapses subgraphs with similar motifs into single nodes to provide an overview of biomedical workflow-based provenance graphs. Maguire et al.~\cite{maguire2013visual} design pictograms for encoding motifs in biological workflows. These workflows exhibit behavior similar to fork-join parallelism, which does not match our communication patterns. More generally, Dunne and Shneiderman~\cite{Dunne2013Motifs} simplify node-link diagrams by replacing clique and fan subgraphs with a small representative solid shape. While these approaches compress heavily, often to icon-scale, our circumstance allow us more space, though not enough to draw the original encoding in a discernable way. Seeking to retain the visceral notion of each structure, we design depictions to evoke the larger pattern, based on how they are interpreted by viewers. 

\section{Preliminary Study}
\label{sec:prelim}

To aid the design of scalable representations of communication patterns and design of experiments for evaluating both new and existing representations, we executed a preliminary study to explore how people interpret communication in Gantt charts. This study focused on how communication is presently drawn and sought to understand what visual factors are considered when differentiating patterns. 

There are two ways people can view communication lines in Gantt charts. The first is to look at all PEs. We call this the {\em full} representation. The problem is when the chart has many PEs which communicate, i.e., it is {\em dense}, the lines may overlap to the point of a solid shape. The other way is to look at a subset of the PEs (rows). We call this the {\em partial} representation. The problem here is that key indicators about the pattern, such as grouping, may not appear in a given window.

Prior to the study, we identified several factors that might be used in interpreting these views: the density of a chart, the structure being shown, the grouping of a structure, the representation type ({\em  partial} or {\em full}), and the stride of a structure. A study across the full range of these factors would be unreasonably large, so we designed a qualitative study to limit our focus, following the ``factor mining'' evaluation pattern discussed by Elmqvist and Yi~\cite{elmqvist2015patterns}.

\begin{figure}
    \centering
    \includegraphics[width=\columnwidth]{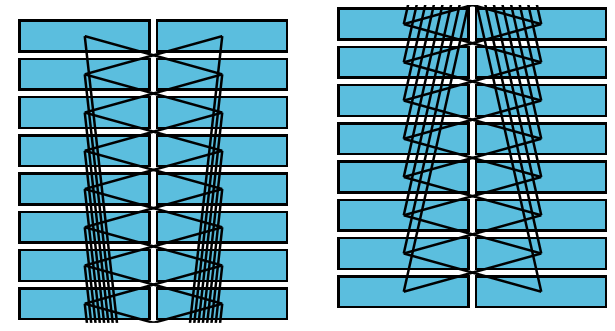}
    \caption{Example image prompt from our preliminary study. These are both {\em partial} representations of a stencil pattern, emulating being zoomed-in on a Gantt chart.}
    \label{fig:interview_prompt}
\end{figure}

Our procedure was to interview participants while showing them paired images of communication patterns in idealized unit time Gantt charts. We decided on semi-structured interviews to allow probing and elaboration of ideas. The paired images varied in number of rows, representation type, and whether they were the same pattern. \autoref{fig:interview_prompt} shows an example prompt. For each prompt, we asked three questions:
\vspace{0em}
\begin{enumerate}
    \itemsep=0em
    \item Please describe the pattern of lines on the right.
    \item Please describe the pattern of lines on the left.
    \item Do you think that these two patterns are the same? Why or why not?
\end{enumerate}

We recruited seven participants, one from our university and six at the Supercomputing 2019 conference. Six reported computing experience and two reported HPC experience. Of those two, one had analyzed HPC performance. None had prior experience with trace visualization. 

The interviews were recorded, transcribed, and coded for common themes, resulting in over 170 unique codes. We describe the most frequently used. See the supplemental material for a full list of codes.

\texttt{Line angle} was used by all participants to describe patterns of lines and justify comparisons between them, with three referencing it over 10 times. \texttt{Line direction} was mentioned by all but one participant, but mostly to describe a pattern rather than compare. Although a line direction could be the result of an angle, it was coded distinct from \texttt{angle} since respondents would describe lines as going ``up," ``down," ``left," or ``right". Their intent was distinct from when they mentioned angle. A few participants mistook the multiple wrap-around lines for \texttt{one line} in ring patterns, even on repeated prompting.

Four participants made note of the \texttt{background}---the boxes and columns the lines connected. This code was also used both to describe an individual pattern and differentiate between pairs.

\texttt{Comparison} and \texttt{uncertainty} were co-occurring codes. Participants rarely felt confident comparing charts. We surmise this uncertainty comes from the fact that simple changes of height and representation alter visual factors they relied upon. Four participants discussed \texttt{transformation} during comparisons, describing how one depiction could be ``rotated'',  ``squashed/stretched'', or ``enlarged.''

Participants had the most difficulty with prompts depicting stencil patterns. They were uncertain of line extents and misidentified discrete lines as single lines or vice versa. Of the four interviewees exposed to stencil patterns, three found them viscerally off-putting, pausing with surprise when presented with them and one calling them ``a mess." 

\vspace{1ex}

\textbf{Discussion.} Participants generally relied on features such as line angle, line direction, and background markings in interpreting patterns. We expect all three factors to be harder to interpret with large numbers of PEs as many angles map to the same pixels and background markers are aggregated or removed.

The line factors are functions of the pattern stride and the height, and therefore density, of the chart, suggesting density significantly impacts recognition. This is further supported with participants' difficulty in recognizing line separation at severe angles, such as the wrap-around lines in rings.

The impetus of participants to identify patterns as similar under a stretch factor matches how the patterns are extensible to PE count. However, communication patterns are not extensible under rotation or inversion, as some discussed. We use this finding to inform tutorial material in subsequent study designs.

Participants reactions to stencils suggest that the idealized unit time depiction is too complex even at small scales. As we want to understand how people can interpret patterns as they scale up, we conclude that stencils are not appropriate for early work towards this goal as they are too difficult to interpret at small scale.

Based on the results of this preliminary study, we decided to take line angle and background into careful consideration for proposed designs. We ultimately fixed line angle at set values and obscure background features to encourage focus on pattern types over strides and to maintain discernability. We also decided to remove stencils from our pattern type factor and randomize across our other factors in further studies.

\section{Scalable Representations for Communication Patterns}
\label{sec:design}

Using the lessons learned in our preliminary studies, along with existing case studies~\cite{isaacs2014combing, Isaacs2016} using idealized unit time, we design a set of strategies for representing communication patterns when there are too many PEs to draw distinct communication lines in Gantt charts. We first describe our design goals. Then, we present our designs. Finally, we discuss initial feedback from experts familiar trace analysis in HPC.

\subsection{Design Goals}

Our goal is to design a representation of communication in execution traces that (1) aids users in recognizing and understanding what communication is occurring in that temporal and logical position in the Gantt chart and (2) is agnostic to the number of processing elements, thereby scaling to larger traces. These goals are derived from usage and scalability limitations noted in prior work~\cite{isaacs2014combing}. 

We limit our focus to scaling in PEs (y-axis) rather than time. Traces are typically explored using a time window, so we focus on that case. Adapting a design or creating a new one for compressed time settings we leave for future work.

Based on our preliminary study (\autoref{sec:prelim}), we chose to focus on offset, ring, and exchange pattern types as stencils require more design consideration even at small scales. 

\subsection{Visualization Design}

Our design process began with open brainstorming on paper, which we include in the supplemental material. We tried a variety of strategies, including linked views and added channels to the traditional Gantt chart encoding rules. However, most of these retained scaling problems, leading us to focus on designs centering on glyphs.

In designing the representation, we considered the saliency of what was to be encoded (e.g., temporal range, pattern type, grouping, stride) and efficacy of available channels, taking into account that the design needs to be incorporated in a Gantt chart. For example, temporal range is set to a horizontal position matching where a pattern would be drawn in a full chart. See supplemental materials for a table containing discussion of channel considerations.

\begin{figure*}
    \centering
    \begin{subfigure}{0.18\textwidth}
         \centering
         \includegraphics[width=\textwidth]{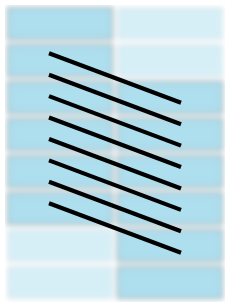}
         \caption{Continuous offset pattern}
         \label{fig:noc}
    \end{subfigure}
    \begin{subfigure}{0.18\textwidth}
         \centering
         \includegraphics[width=\textwidth]{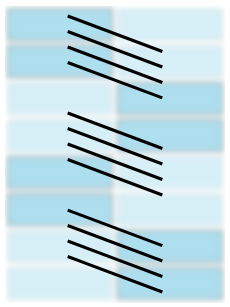}
         \caption{Grouped offset pattern}
         \label{fig:nog}
    \end{subfigure}
    \begin{subfigure}{0.18\textwidth}
         \centering
         \includegraphics[width=\textwidth]{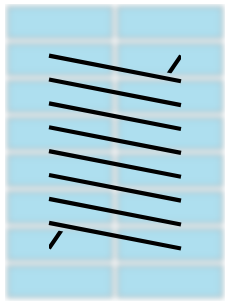}
         \caption{Continuous ring pattern}
         \label{fig:nrc}
    \end{subfigure}
    \begin{subfigure}{0.18\textwidth}
         \centering
         \includegraphics[width=\textwidth]{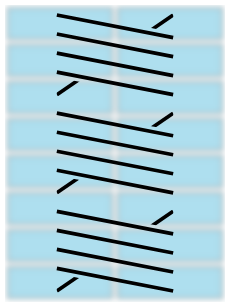}
         \caption{Grouped ring pattern}
         \label{fig:nrg}
    \end{subfigure}
    \begin{subfigure}{0.18\textwidth}
         \centering
         \includegraphics[width=\textwidth]{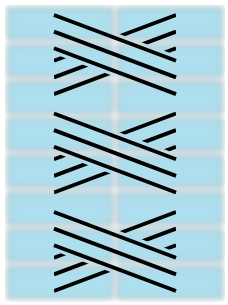}
         \caption{Exchange pattern}
         \label{fig:neg}
    \end{subfigure}
    \caption{Examples of our designs for five communication patterns. They are reminiscent of the underlying communication pattern encoding, but not aligned to the underlying chart and agnostic to the number of rows the underlying pattern repeats over. 
    Grouped representations fill the vertical space to indicate that the repetition continues from the top of row to the bottom.}
    \label{fig:abstract_designs}
\end{figure*}

We prioritize the type of pattern before the grouping factor or stride. The rationale is that the pattern type is fixed by the source code while the grouping and stride are often computed from the problem size and number of resources. Therefore, a user will recognize pattern type first before considering other factors. \autoref{fig:abstract_designs} shows the resulting designs.

\vspace{1ex}

\textbf{Encoding Pattern Type.} To encode the pattern type, we started with the overall shape of of the pattern when drawn at small scale with a small stride. Offsets are drawn with angled repeating lines forming a rhombus-like shape. We use a fixed distance between lines and draw as many will fit in the relevant area.

Rings add indicators of the ``wrap-around'' communication. However, unlike fully drawn rings, we only render the protruding segments at the ends of the shape. There are two main rationales for only drawing protruding segments: (1) we want to indicate this is an abstraction and (2) participants in our interviews found the crossing lines difficult to disambiguate. The number of protruding segments is proportional to the stride of a ring. 

Exchange patterns are drawn as a series of symmetrical ``x'' shapes and avoid direct crossings for the same reason as rings. The number of lines in each cross is proportional to stride of the exchange. Short stride exchanges will exchange between only a few PEs, a long stride exchange spans many PEs. Our glyphs approximate this by increasing the number of crossing lines as stride increases.

\vspace{1ex}

\textbf{Encoding Grouping Factor.} To represent grouping, we partition the available area vertically and repeat the pattern type drawing in those partitions. More formally, the encoding rule to show ``grouping" is repetition and alignment on a non-common scale. The number of partitions is determined by the available vertical space in a chart.

\vspace{1ex}

\textbf{Encoding Stride.} We express the notion of stride through the angle of lines used in our pattern types. As people had difficulty with steeply angled lines in our preliminary study, we limit the angles to a range of 15 degrees to 60 degrees. Therefore, these do not match the encoding of a full view. Instead, they hint at the magnitude of distance over which communication is occurring. This allows users to see that there are differences in stride between glyphs, but not necessarily calculate the exact stride visually.

\vspace{1ex}

\textbf{Temporal range.} Rather than show the exact range, we place the glyphs on the x-axis so they are centered in their range. If two structures overlap, they are placed alongside one another. 

\vspace{1ex}

\textbf{Incorporation in Gantt Charts.} These are designed to be used in Gantt charts when exact lines would be too dense to be interpreted. The underlying interval rectangles will still be drawn. The color encoding of these intervals was shown to be a secondary indicator in our preliminary study, so we preserve them. We add a slight blur effect to the background as another signifier that the glyphs are an abstraction and should not be confused for exact lines.

\subsection{Expert Feedback}
\label{sec:expertfeedback}

We sent our designs to two HPC experts for feedback regarding both the designs themselves and the overall approach. Both experts were familiar with idealized unit time representations of traces. The first expert, E1, had previously collaborated on this strategy with the authors but was not involved in any of the work presented here. The second expert, E2, had managed an integration of the strategy into an HPC center's performance tools, referencing the open-source research code~\cite{isaacs2014combing} but using an alternate calculation method and front-end technology.

We sent both experts a short email with a PDF describing the visualizations with comparisons to fully drawn traces and showing how they might be applied in practice, including a few complicated examples such as zoomed-out time and idle processes. (See supplemental materials.) We asked if and how the strategy would be useful and if there were any suggestions or concerns. E1 responded the designs ``definitely look helpful,'' noted the trade off in exactness, and then pointed out figures which led to ambiguities in his view. He also identified a error where the mock-up did not match the underlying trace. E2 noted that stride is less important and wondered how the translation from data to glyph would be calculated. He suggested the strategy might also be helpful for collective communications (e.g., broadcasts, all-to-all, reductions), a set of patterns we did not consider in this work.

We interpreted these responses to suggest the designs were worth further study, particularly E1's ability to interpret well enough to detect an error and E2's interest in further patterns. However, there are design decisions in applying these glyphs in some scenarios, particularly in zoomed-out time, that require refinement. We leave these cases for future iterations and instead focus on how the base designs could be interpreted by a wider range of users in a controlled study.

\section{Experiment Methodology}
\label{sec:methodology}

We evaluate the efficacy of the proposed design and the existing methods through a pre-registered~\footnote{
https://osf.io/zwrgk}
user study. The final design was refined through four pilot studies, two in-person task-design-focused pilots and two online holistic pilots. The study source code and information about the pilots is available in the supplemental materials.

\subsection{Experiment Overview}

We design a within-subjects study with one factor: representation type. This factor has three levels: {\em full}, {\em partial}, and {\em new}, where {\em full} and {\em partial} are the zoomed-out and zoomed-in versions of the existing Gantt chart and {\em new} is the design described in \autoref{sec:design}.

We measure four dependant variables: (1) total accuracy across all question types, (2) accuracy for pattern type classification questions, (3) accuracy for grouping classification questions, and (4) accuracy for stride differentiation questions. We measure accuracy as the ratio of correct answers / total answers, per participant. We chose to make the denominator total answers instead of \textit{total questions} to account for missing values in the dataset, e.g., where the platform failed to log a user's response properly. 

In classification questions, participants were shown a chart and asked which of the three pattern types it represented and then separately, whether the pattern was continuous or grouped. In discrimination questions, participants were shown two charts and asked if they exhibited the same stride. Participants were shown eight target patterns, rendered in each of the three styles, with two categorization questions (type, grouping) and two discrimination questions (both stride comparison against two full or two partial charts), resulting in (8 charts x 3 conditions x 2 categorizations) $+$ (8 charts x 3 conditions x 2 discriminations) $=$ 96 total trials. Five of these patterns were $>1000$ rows with each set to one  of the five configurations of structure and grouping we support with our new designs (\autoref{fig:abstract_designs}).
    
Chart density, pattern strides, and pattern grouping are randomized to increase generalizability across specific pattern instances and following our findings in \autoref{sec:prelim}. Chart density was varied between a binary selection of 1280 or 2560 PEs, to simulate dense charts common in reasonably sized parallel programs. We did not scale these charts to still-realistic sizes of 10,000 PEs as they were even less discernible and conclusions derived from these ``smaller'' charts would extend to even denser ones.  Pattern stride varies in the range of 2 to 10. Stride is kept relatively small in our pattern generation to accommodate possible configurations of grouped patterns which have relatively small grouping factors. However, as we discuss in \autoref{sec:analysisdiscussion}, this choice may have given {\em partial} representations an advantage. Patterns were grouped into bundles of 4 to 16 lines because they would be possible grouping configurations at low or high numbers of PEs.

We did not test the placement of the representations within a larger Gantt chart. Our rationale is that we first seek to understand efficacy in a simple isolated case so we can draw conclusions for further design.

\subsection{Procedure}
    The study begins with an informed consent document and short questionnaire regarding demographic information, including prior experience with programming, HPC, and performance analysis. There are then two experimental modules: classification and discrimination. Each module begins with a written tutorial explaining the context, seven practice questions with answers and explanations shown, and then the set of 48 recorded trials, one per screen. During the trials, a progress bar is shown. After both experimental modules, participants were asked which representation they preferred and given a free-form text fields to elaborate on their opinions of the experiment and tutorial.

    The classification tutorial introduces participants to the concepts of parallel resources, tracing, Gantt charts, and communication patterns therein, with emphasis on pattern types and grouping. The discrimination tutorial reinforces the concept of pattern stride. We divided questions into these two modules with separate tutorials following concerns about information overload in our first online pilot.
    Participant feedback in the second online pilot noted the tutorials were complex and long, but also went into appropriate depth. All participants rated them as ``moderately helpful'' (5), ``helpful'' (2), or ``very helpful'' (3).

\subsection{Task Design}

    We asked users to perform two types tasks:
    \vspace{-.5em}
    \begin{enumerate}[start=1, label={\bfseries T\arabic*}]
        \itemsep0em
        \item Classify a pattern in a chart
        \item Discriminate stride between two charts
    \end{enumerate}
    
\begin{figure}[htb]
        \centering
        \includegraphics[width=\columnwidth]{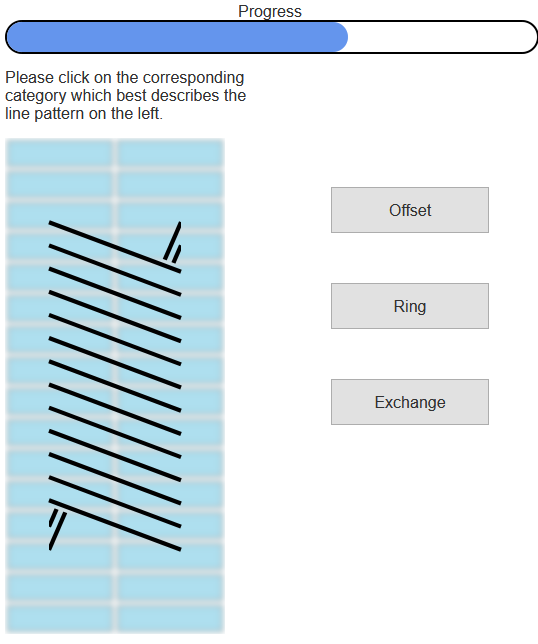}
        \caption{A pattern type classification question. Users are prompted with a chart on the left side of the screen and given a 3-alternative forced choice prompt to select the underlying communication structure.}
        \label{fig:class-pattern}
    \end{figure}

\begin{figure}[htb]
        \centering
        \includegraphics[width=\columnwidth]{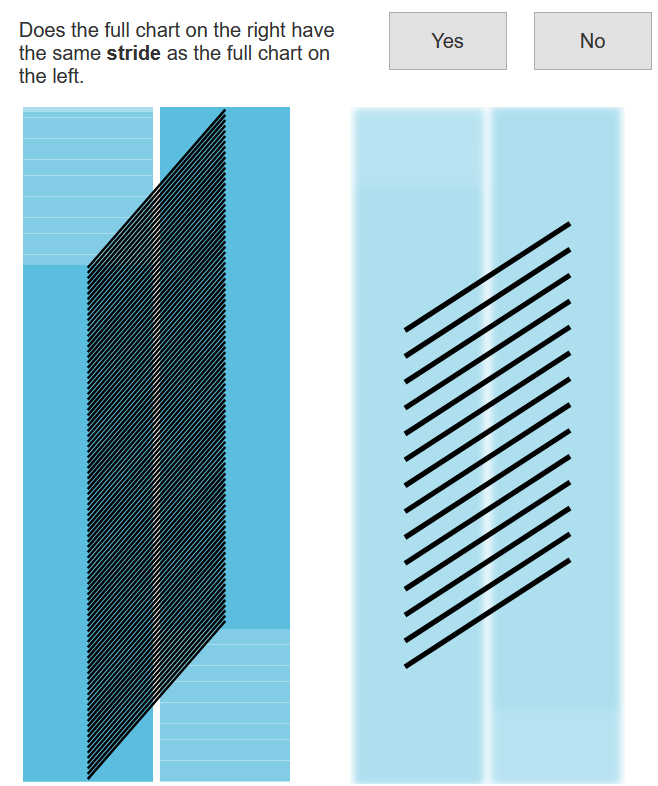}
        \caption{An example of a stride discrimination question. Users are prompted with two charts, of different representation or number of rows and asked if they have the same stride.}
        \label{fig:discrim-ex}
    \end{figure}

    The classification task was broken into two separate questions, a three-alternative forced choice for the pattern type and a two-alternative forced-choice between grouped or continuous. These tasks were not necessarily asked in sequence. \autoref{fig:class-pattern} shows an example pattern type classification question. The discrimination task asks participants if two representations have the same stride. \autoref{fig:discrim-ex} shows an example.
    
For both types of tasks, participants were instructed to answer quickly with a suggestion to take less than 15 seconds per question. The rationale is that the use cases driving these questions are similarly quick, aiding the user in directing focus for further analysis. In practice, users responded much quicker with a mean response time of 4.3s and median response time of 3.1s. 
    
    The tasks were designed iteratively through piloting. We decided to separate the pattern attributes of type, grouping, and stride following the first offline pilot and the preliminary study which both asked a more general discrimination task: if the charts had the ``same pattern.'' We found it difficult to disambiguate errors in these pilots, especially as participants took different meanings of the word `pattern.'
    
    We initially chose discrimination tasks based on the scenario that performance analysts might want to identify which regions in a Gantt chart are iterations of the same communication operation or even the same code. We switched pattern type and grouping to classification under the rationale that this identification was necessary to assess same-ness and that recognition tasks for these higher level abstractions are important. We kept stride as a discrimination task because the exact number is less important and as stated previously, dependent on several execution-time calculations.
    
    We also considered a drawing task, where participants would be asked to copy a pattern from one representation onto a blank background of a different size, as a holistic test. We ran a crayon and paper pilot of this approach but found the responses difficult to interpret, even with follow-up interviews.

\subsection{Participants}
    Ideally the participants in this experiment would all be familiar with HPC, a variety of HPC applications, HPC traces, Gantt charts and logical time. However, we previously struggled to find sufficient numbers of people with this expertise to participate in a quantitative study. As our preliminary study did not suggest differences in people based on their relative HPC experience, we decided to widen the pool to people familiar with computing in general. Our rationale was that these people could serve as a proxy for beginners in visual trace analysis, similar to the HPC workers we previously interviewed. Furthermore, they could understand the context given the tutorial.
    
    We arrived at a target of 35 participants based on our second online pilot which had 10 participants and was run on the target platform. Using the means and variances for each hypothesis from that pilot, we calculated the participants necessary to achieve a power of $1 - \beta=0.8$ with our target significance for each hypothesis and took the maximum participants required across all hypotheses.

    Using the Prolific~\cite{prolific} crowd sourcing platform, we recruited 43 participants who reported at least ``beginner'' level coding experience in their profile and were using desktop or laptop machines. Participants were payed a flat rate of \$9.00 for our estimated 35 minute experiment, reflecting a target wage of \$15/hr.
    
    We included four trivial questions, two in each module, to serve as guards. We rejected responses that failed one or more of these questions and collected data until we had 35 valid responses. We rejected eight responses based on these guards, of which five had total response times of less than half of the estimated experiment time. Of the valid responses, participants reported their ages in [20,56] with the median being 26. Twenty-eight reported as men and seven as women.

\subsection{Experiment Platform}
    The experiment platform is a custom built website hosted on a Heroku web application server~\cite{experiment_platform}. The website is composed of a Flask~\cite{flask} (Python) server and a Javascript front-end. All charts were drawn with D3js~\cite{d3js}. The source code is included in the supplemental materials.
    
    Participants were restricted through the Prolific platform to desktop or laptop devices. Screen resolutions ranged from 1093 $\times$ 615 to 2560 $\times$ 1440, with a mode of 1920 $\times$ 1080.

\subsection{Hypotheses}

  Based on data from the second online pilot as well as our own design goals, we arrived at the following hypotheses for experiment:
    
    \vspace{-.5em}
    \begin{enumerate}[start=1, label={\bfseries H\arabic*}]
        \itemsep-.3em
        \item New designs will be approximately the same in accuracy as traditional full representations for identification of structure, grouping and stride when taken all-together.
        \item New designs will enable more accurate identification of communication trace structure, for sufficiently dense charts ($>1000$ rows/processes), than either partial or full representations.
        \item New designs will enable more accurate identification of communication trace structure "grouping", for sufficiently dense charts ($>1000$ rows/processes), than either partial or full representations.
        \item New designs will enable less accurate identification of communication trace pattern stride than either partial or full representations.
        \item Partial or "zoomed in" representations will result in less accurate identification of communication trace pattern stride, in addition to grouping and pattern identification than either full representations or new designs.
    \end{enumerate}

In designing the {\em new} representations, we chose to trade off exactness in stride for discriminability in pattern type and grouping. Thus, we expected the {\em new} representations to result in increased accuracy for pattern type (``structure'') and grouping (H2, H3), but worse accuracy for stride (H4). Based on pilot data, we expected these trade off to ``even out'' in overall accuracy between {\em new} and {\em full} (H1), with {\em partial} representations lagging based on pilot data (H5).

\section{Results and Analysis}
\label{sec:analysis}
    We evaluate the hypotheses outlined in \autoref{sec:methodology} using an Analysis of Variance test (ANVOA) with a Tukey HSD Post-Hoc test to evaluate the impact of and pairwise significance of our independent variables. We choose a significance of $0.05$ and apply a Bonferroni correction for multiple testing, resulting in an effective significance of $0.01$. 
    
    To justify the validity of using ANOVA, we applied a Shapiro-Wilk test to evaluate the normality of each subset of data investigated in our independent variable. For each level, the Shapiro-Wilk p-value exceeded the target $0.05$, so we proceeded with the test. We further supplement the analysis with bootstrap confidence intervals (95\%, 1000 trials) which are consistent with the ANOVA results.
    
    In our analysis of pattern classification tasks, we examine three confusion matrices in an exploratory fashion for further insight into how users may have arrived at the incorrect answers. Although 8 questions were asked of participants, three were included for a potential exploratory analysis and were not sufficiently dense by our definition of $>1000$ rows and were not included in the following hypothesis testing.
            
    \begin{figure}
        \centering
        \includegraphics[width=\columnwidth]{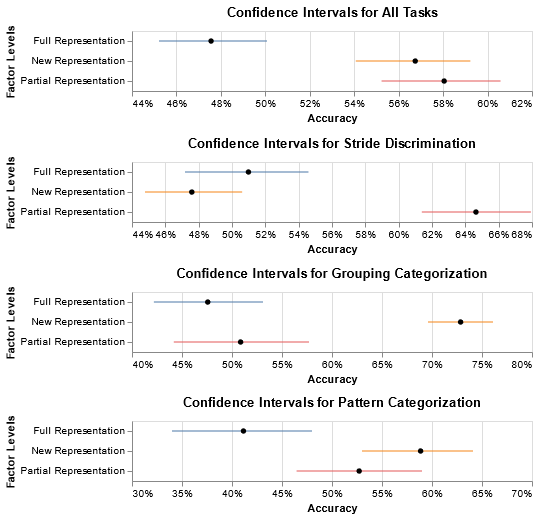}
        \caption{95\% Bootstrap confidence intervals. }
        \label{fig:all_ci}
    \end{figure}
    
    \vspace{1ex}
    
\textbf{H1 \& H5: {\em New} and {\em Partial} outperform {\em Full} in Overall Accuracy.} H1, that the {\em full} and {\em new} representations would have the same overall accuracy, and H5, that participants would be less accurate overall with {\em partial} representations, are not supported by the results. \autoref{fig:all_ci} (top) shows the bootstrap confidence intervals of accuracies aggregated across all tasks: pattern identification, grouping identification, and stride comparison. In the confidence intervals we see that there is no overlap between {\em new} and {\em full} representations as hypothesized. The overall effect of this factor on accuracy is statistically significant  (F$_{2,32}$ = 17.58, p $<$ .01), and the effect difference between {\em new} and {\em full} is determined to be statistically significant (p $<$ .01).

\begin{figure*}
    \centering
    \begin{subfigure}{0.28\textwidth}
        \centering
        \includegraphics[width=\textwidth]{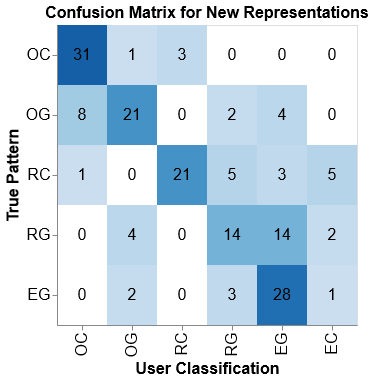}
        \caption{Confusion matrix for {\em new} representations.}
        \label{fig:abstract_conf_mat}
    \end{subfigure}
    \begin{subfigure}{0.28\textwidth}
        \centering
        \includegraphics[width=\textwidth]{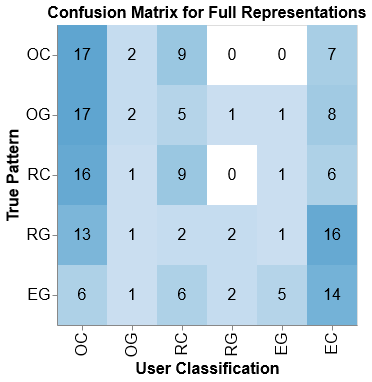}
        \caption{Confusion matrix for {\em full} representations. }
        \label{fig:full_conf_mat}
    \end{subfigure}
    \begin{subfigure}{0.37\textwidth}
        \centering
        \includegraphics[width=\textwidth]{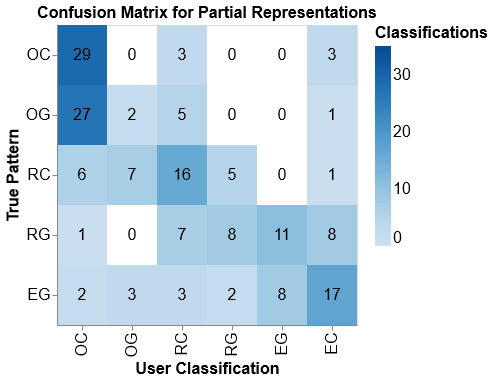}
        \caption{Confusion matrix for {\em partial} representations.}
        \label{fig:partial_conf_mat}
    \end{subfigure}
    
    \caption{Confusion matrices showing how participants classified trial patterns in our study. To preserve space, the pattern classifications are abbreviated: O(ffset), R(ing), E(exchange), C(ontinuous), G(rouped). The matrix has an extra column because participants were able to answer that a pattern was ``Exchange Continuous" but no such patterns existed in our question set. }
    \label{fig:confusion_matrices}
\end{figure*}

The {\em new} representations performed better than {\em full} when all tasks were considered. In later analyses we will see this is driven almost entirely by the classification questions, despite poor performance with discrimination questions. Unexpectedly, the {\em partial} representations also performed better than {\em full} with no significant difference to the {\em new} representations.

\vspace{1ex}

\textbf{H2: New Outperforms Full in Pattern Classification Accuracy.} H2, that classifications would be more accurate with the {\em new} designs for dense charts ($>1000$ PEs), is partially supported by our results. The effect of representation type on pattern classification accuracy was determined to be statistically significant (F$_{2,32}$ = 7.62, p $<$ .01). The Tukey HSD test indicated a significant difference between {\em new} and {\em full} representations (p $<$ .01), but not {\em new} and {\em partial}. \autoref{fig:all_ci} shows the bootstrap confidence intervals.

    Though the {\em new} representations performed better than the existing ones, absolute accuracy does not exceed an average of 59\% with a maximum of 64\% and a minimum of 53\%. To further understand this behavior, we examine a confusion matrix (\autoref{fig:abstract_conf_mat}). The majority of errors are Ring-Grouped (RG) patterns being misidentified as Exchange-Grouped (EG). These depictions are visually similar. As noted previously, an exchange is identical to a ring with a stride of half the participating PEs, so participants may have had trouble differentiating when the depicted stride was close to half, which is more likely in grouped scenarios with rings which have fewer lines in the glyph.
    
    \autoref{fig:full_conf_mat} shows the confusion matrix for the {\em full} trials. Most errors were due to user selecting Offset-Continuous (OC) when the pattern was something else. It is unclear if this was due to signifiers like spacing between groups and wrap-around lines being obscured by the density or users selecting the first option when they were unsure. A second feature to note is that Ring-Grouped (RG) and Exchanged-Grouped (EG) charts are commonly mis-classified as Exchange-Continuous (EC). As noted before, these patterns can be quite similar, both in structure and appearance.
    
    In \autoref{fig:partial_conf_mat}, mis-classifications are most densely clustered around Offset-Grouped (OG) as Offset-Continuous (OC) and Exchange-Grouped (EG) as Exchange-Continuous (EC). These errors are likely attributable to the fact that a ``break" signfiying a grouping in a pattern can occur past the top or bottom edge of a view. If the break cannot be seen, determining grouping is just a guess. Furthermore, we see that Ring-Grouped (RG) are mis-classified as EG and EC for the same reasons noted above: they are both visually and structurally very similar.

        \vspace{1ex}

\textbf{H3: {\em New} Outperforms {\em Full} and {\em Partial} in Grouping Accuracy.} H3, that participants would be more accurate with {\em new} representation than {\em full} or {\em partial} for identifying a pattern as grouped or continuous, is supported by the results. The effect of representation type on classification accuracy for ``grouping" questions is statistically significant (F$_{2,32}$ = 23.34, p $<$ .01) with the Tukey HSD tests indicating pairwise significant between {\em new}-{\em full} and {\em new}-{\em partial}. The {\em new} representation had a median accuracy of 73\%, while {\em full} and {\em partial} accuracies are within range of random guessing. \autoref{fig:all_ci} shows the corresponding bootstrap confidence intervals.

\vspace{1ex}

\textbf{H4: {\em Partial} Outperforms {\em New} and {\em Full} in Stride Estimation.} H4, that participants would be less accurate with the {\em new} representations when it came to stride estimation, is partially supported by our results. A significant difference is shown in the ANOVA (F$_{2,32}$ = 28.71, p $<$ .01), with the Tukey HSD showing a significant pairwise difference between {\em partial}-{\em new} and {\em partial}-{\em full}, but not {\em new}-{\em full}. \autoref{fig:all_ci} shows the bootstrap confidence intervals.
  
    As expected, {\em new} did not perform well on stride comparisons as the encoding is only approximate. Though the encoding is direct for {\em full} representation, the resulting visualization can be ambiguous when several strides map to the same pixel. {\em Partial} representations can contain the stride, if small enough. This may explain the low accuracy ($<65$\%) even in {\em partial} representations.
    
    \vspace{1ex}

\textbf{Post-trial Survey.} In the post-trial survey, 18 (51.4\%) of participants preferred the {\em new} representation, 14 (40.0\%) preferred the {\em partial} representation, and 3 (8.6\%) preferred the {\em full} representation. Only two commented on the designs, one stating that {\em full} was ``hard to discern'' and the other stating {\em full} looked ``the most practical''.

\subsection{Discussion}
\label{sec:analysisdiscussion}

The study results show that  {\em new} and  {\em partial} representations are strongest in overall accuracy across pattern type, grouping, and stride. Breaking this down, participants were more accurate with then  {\em new} representation for detecting grouping versus continuous patterns and the  {\em partial}  representation for detecting equivalent strides. While there was no statistically significant difference between  {\em new} and  {\em partial} in terms of pattern type, the median for  {\em new} was higher and most of the errors were attributable to patterns that are very similar to each other.

These results suggest the  {\em new} representation is helpful in what it was designed to do---provide a scalable overview of the communication pattern, emphasizing the type and grouping of a pattern first. The visualization de-prioritizes strides by design, a decision reinforced by the feedback from expert E2.

We were surprised how well the  {\em partial} representation performed overall, given the results of our pilot studies. This view occurs when zooming-in to the Gantt chart, suggesting that further emphasis be given in developing methods that take advantage of this view, such as lenses, or systems that support low latency rendering for interactive navigation of large scale traces.

One caveat to the performance of  {\em partial} representations is that our trials only included small stride values of ten or less, meaning many of them had relatively gentle angles and may have fit within the  {\em partial} view. Larger strides would have incurred steeper angles, something participants in our preliminary study struggled with in rings. 

Given the complementary strengths of the  {\em new} and  {\em partial} representations, designing an interactive Gantt chart to select between the two based on zoom-level may aid users in understanding communication.

Participant preference mostly tracked with participant performance, with the majority of participants preferring the {\em new} designs but a sizable minority preferring the {\em partial} views. This further suggests a complementary approach. Furthermore, the gap in preference to {\em full} suggests the {\em new} designs enhance satisfaction in addition to boosting accuracy in the core use case they were developed for.

Lastly, we note that the accuracy levels were not particularly high across the board, with the highest being recognizing grouped versus continuous patterns with the  {\em new} representation at 73\%. This low accuracy suggests that the present methods,  {\em full} and  {\em partial}, do not serve people well and while the  {\em new} representation improves the situation in some cases, there is still a gap in interpreting communication.
    
\subsection{Limitations and Threats to Validity}

The participants in this study were not HPC experts and did not have experience with large-scale parallel programs, the communication patterns they use, or large-scale trace visualization. The results with an expert population might differ. However, we did not think we could recruit such a population of sufficient size and even people who visualize traces regularly may not be familiar with idealized unit time as it is not yet supported by commercial trace visualization software.

The categorization task in this experiment involved three types of patterns and a binary choice between grouped and continuous. In a realistic setting, participants need to recognize patterns without constraints.

This experiment compared accuracy of static visualizations with relatively short interpretation periods to simulate browsing behavior. Results may differ in more focused analysis when a user engages with the system, panning and zooming data of interest. 

All representations were presented to participants in isolation. However, the ultimate use case is for them to be presented in the context of an interactive Gantt chart with potentially multiple patterns at the same time. This additional context could affect the strategies people take and their resulting accuracy.

\section{Reflections and Lessons Learned}
\label{sec:reflections}

We reflect on the concept of a pattern, communication recognition in Gantt charts, designing with differing participant pools, and designing in difficult-to-implement scenarios.

\vspace{1ex}

\textbf{`Pattern' is an ambiguous concept, but the domain can help ground and decompose the concept for design.} We use the term `communication pattern' because that is the term our domain expert collaborators used in previous projects. Our goal was to understand factors in interpreting these patterns and how to use them to create a scalable visual representation. However, the term {\em pattern} has many different interpretations. 

In our preliminary study, we observed that participants did not have a concrete understanding of the idea of a pattern or of the constraints the domain had on them. They discussed operations for transforming between instances, some of which could be valid in context, such as stretching and squishing, and some which could not, such as rotation. In effect, the preliminary study acted as an intervention~\cite{Bigelow2021}, correcting our assumptions on how we and others interpreted the notion of `pattern.' On reflection, we have considered alternate terms like `class,' `family,' `motif,' and `extensible structure,' but we suspect they would have the same ambiguity.

As communication patterns have an extensible, repetitive structure to them, we initially considered them similar to classes of graph structures in graph theory. However, communication patterns are not the same under rotational symmetry, something we realized was a potential factor due to the preliminary study. 

Understanding what factors participants were considering helped us in two key ways. First, it helped us to dissect the notion of `communication pattern' into three factors: type (base structure), grouping, and stride. We then assigned a priority to these factors, based on domain use cases, to drive the design. 

Second, the preliminary study revealed the visual strategies people preferred, such as line angle. In the context of the domain, this preference shifted focus on a less important factor, so we designed our solution to minimize its saliency.

\vspace{1ex}

\textbf{Present strategies for visualizing communication in Gantt charts are not sufficient.} {\em Full} representations, what is shown when Gantt charts show all rows, only narrowly outperformed random guessing (41\% vs. 33\%) in our study. Though our new designs improve accuracy versus full overviews, even in this constrained scenario, there is significant room for improvement. The performance of {\em partial} representations suggests improving access to them, for example, with lenses or smooth navigation, may also improve understanding of structures.

\vspace{1ex}

\textbf{Participants with computing experience served as a proxy for beginners in trace analysis.} Our desire to improve visual scalability in Gantt charts was motivated and informed by visualization outcomes of a long-term collaboration with domain experts. However, relying only on those experts, who are most familiar with the visualization, as participants in this next step could bias the results. Another practical issue is the difficulty of recruiting more domain experts due to demands on the time of people with this expertise. 

Following the strategy of McKenna et al.~\cite{McKenna2016}, we sought to use multiple pools of people in our design. In particular, we recruited people with some HPC familiarity in our preliminary study, people with high trace visualization familiarity for informal design feedback, and people with general computing experience for our controlled study, which required the most participants.

Our preliminary study did not suggest a noticeable effect of relative experience of people in HPC who did not have trace visualization experience. Thus, we recruited people with more general computing experience for the larger study, assuming they would be similar to an HPC worker who was new to analyzing traces visually or exposed to a new method of analyzing traces. Given that participant recruitment is a perennial problem, understanding differences in adjacent populations such as these, could aid practical aspects of study design. 

\vspace{1ex}

\textbf{There are scenarios encouraging speculative visual design.} The idealized unit time approach used in Ravel required complicated domain-aware algorithms to compute~\cite{Isaacs2015,Isaacs2016}. Suspecting identifying communication patterns might be similar, we chose to investigate the efficacy of the proposed visualization first, to determine if investing in the development domain-aware algorithms would be prudent. Our rationale was that even if a perfect algorithm could not be achieved, what we learn in the study could be applied to an imperfect one. 

The patterns we investigated could also be annotated in the source code. We learned a similar strategy was taken by E2 (\autoref{sec:expertfeedback}) for the Ravel-type visualization. E2 saw the visualization and was motivated to engineer a solution without the more intensive algorithm. This demonstrates the utility of speculative visualizations such as these.

\section{Conclusion}
\label{sec:conclusion}

We presented designs and evaluations for representing communication patterns, an extensible structure, in Gantt charts as they scale up to thousands of rows, as is common in HPC. Through our preliminary experiment, we observed people's varying ideas on the concept of ``patterns'' as well as visual factors they used to interpret them. These observations led us to decompose communication patterns into three domain-specific indicators and design a visual approach for representing them in a scale-agnostic approach that maintains their ethos. A controlled study with these designs showed they enabled a more accurate identification of salient characteristics compared to large-scale views of communication patterns. This suggests applying our design strategy to other extensible patterns, such as collective communications, may prove effective. In addition to validating our new design, our evaluations confirm our concerns with scalability in Gantt charts, but also suggest further focus on making partial views responsively navigable along with scalable overviews such as ours, could clarify important aspects of communication patterns. 

\section{Acknowledgements}

This work was supported by the National Science Foundation under NSF III-1656958.

\bibliographystyle{abbrv-doi}

\bibliography{main}

\begin{thebibliography}{10}

\bibitem{LULESH}
{H}ydrodynamics {C}hallenge {P}roblem, {L}awrence {L}ivermore {N}ational
  {L}aboratory.
\newblock Technical Report LLNL-TR-490254, Lawrence Livermore National
  Laboratory.

\bibitem{Dosimont2014Ocelotlb}
A spatiotemporal data aggregation technique for performance analysis of
  large-scale execution traces.
\newblock CLUSTER, Sept 2014.

\bibitem{nas}
D.~H. Bailey, E.~Barszcz, J.~T. Barton, D.~S. Browning, R.~L. Carter, and R.~A.
  Fatoohi.
\newblock The {NAS} parallel benchmarks.
\newblock {\em International Journal of Supercomputer Applications},
  5(3):63--73, 1991.

\bibitem{Bigelow2021}
A.~Bigelow, K.~Williams, and K.~E. Isaacs.
\newblock Guidelines for pursuing and revealing data abstractions.
\newblock {\em To appear in IEEE Transactions on Visualization and Computer
  Graphics (Proceedings of InfoVis '20)}, Jan. 2021.

\bibitem{d3js}
M.~Bostock, V.~Ogievetsky, and J.~Heer.
\newblock D3: Data-driven documents.
\newblock {\em IEEE Transactions on Visualization and Computer Graphics},
  17(12):2301--2309, Dec 2011. doi: {{%
10\hspace{.1pt}\discretionary{.}{%
}{.}\hspace{.4pt}1109\discretionary{/}{%
}{/}TVCG\hspace{.1pt}\discretionary{.}{%
}{.}\hspace{.4pt}2011\hspace{.1pt}\discretionary{.}{%
}{.}\hspace{.4pt}185}}


\bibitem{Brendel2016}
R.~Brendel, M.~Heyde, H.~Brunst, T.~Hilbrich, and M.~Weber.
\newblock Edge bundling for visualizing communication behavior.
\newblock In {\em Proceedings of the Third International Workshop on Visual
  Performance Analysis}, VPA, pp. 1--8, 2016.

\bibitem{Cottam2015}
J.~Cottam, B.~Martin, L.~Dalessandro, and A.~Lumsdaine.
\newblock Pixel-oriented techniques for visualizing next-generation hpc
  systems.
\newblock In {\em Proceedings of the 3rd IEEE Working Conference on Software
  Visualization}, VISSOFT, sep 2015.

\bibitem{Zinsight}
W.~De~Pauw and S.~Heisig.
\newblock Zinsight: A visual and analytic environment for exploring large event
  traces.
\newblock In {\em Proceedings of the 5th International Symposium on Software
  Visualization}, SOFTVIS, pp. 143--152. ACM, New York, NY, USA, 2010. doi: {{%
10\hspace{.1pt}\discretionary{.}{%
}{.}\hspace{.4pt}1145\discretionary{/}{%
}{/}1879211\hspace{.1pt}\discretionary{.}{%
}{.}\hspace{.4pt}1879233}}


\bibitem{Dunne2013Motifs}
C.~Dunne and B.~Shneiderman.
\newblock Motif simplification: Improving network visualization readability
  with fan, connector, and clique glyphs.
\newblock In {\em Proceedings of the SIGCHI Conference on Human Factors in
  Computing Systems}, CHI '13, p. 3247–3256. Association for Computing
  Machinery, New York, NY, USA, 2013. doi: {{%
10\hspace{.1pt}\discretionary{.}{%
}{.}\hspace{.4pt}1145\discretionary{/}{%
}{/}2470654\hspace{.1pt}\discretionary{.}{%
}{.}\hspace{.4pt}2466444}}


\bibitem{elmqvist2015patterns}
N.~Elmqvist and J.~S. Yi.
\newblock Patterns for visualization evaluation.
\newblock {\em Information Visualization}, 14(3):250--269, 2015.

\bibitem{libnbc}
T.~Hoefler, A.~Lumsdaine, and W.~Rehm.
\newblock {Implementation and Performance Analysis of Non-Blocking Collective
  Operations for MPI}.
\newblock In {\em Proceedings of the 2007 International Conference on High
  Performance Computing, Networking, Storage and Analysis, SC07}. IEEE Computer
  Society/ACM, Nov. 2007.

\bibitem{Huynh2015DAGViz}
A.~Huynh, D.~Thain, M.~Pericas, and K.~Taura.
\newblock {DAGViz}: A {DAG} visualization tool for analyzing task-parallel
  program traces.
\newblock In {\em 2nd Workshop on Visual Performance Analysis}, Nov. 2015.

\bibitem{Isaacs2015}
K.~E. Isaacs, A.~Bhatele, J.~Lifflander, D.~B\"ohme, T.~Gamblin, M.~Schulz,
  B.~Hamann, and P.-T. Bremer.
\newblock Recovering logical structure from {C}harm++ event traces.
\newblock In {\em Proceedings of the ACM/IEEE International Conference for High
  Performance Computing, Networking, Storage and Analysis}, SC '15, Nov. 2015.
\newblock LLNL-CONF-670046.

\bibitem{isaacs2014combing}
K.~E. Isaacs, P.-T. Bremer, I.~Jusufi, T.~Gamblin, A.~Bhatele, M.~Schulz, and
  B.~Hamann.
\newblock Combing the communication hairball: Visualizing parallel execution
  traces using logical time.
\newblock {\em IEEE transactions on visualization and computer graphics},
  20(12):2349--2358, 2014.

\bibitem{Isaacs2016}
K.~E. Isaacs, T.~Gamblin, A.~Bhatele, M.~Schulz, B.~Hamann, and P.-T. Bremer.
\newblock Ordering traces logically to identify lateness in message-passing
  programs.
\newblock {\em IEEE Transactions on Parallel and Distributed Systems},
  27(3):829--840, Mar. 2016.
\newblock LLNL-JRNL-668754. doi: {{%
10\hspace{.1pt}\discretionary{.}{%
}{.}\hspace{.4pt}1109\discretionary{/}{%
}{/}TPDS\hspace{.1pt}\discretionary{.}{%
}{.}\hspace{.4pt}2015\hspace{.1pt}\discretionary{.}{%
}{.}\hspace{.4pt}2417531}}


\bibitem{isaacs2014state}
K.~E. Isaacs, A.~Gim{\'e}nez, I.~Jusufi, T.~Gamblin, A.~Bhatele, M.~Schulz,
  B.~Hamann, and P.-T. Bremer.
\newblock State of the art of performance visualization.
\newblock In {\em EuroVis (STARs)}, 2014.

\bibitem{Jo2014}
J.~{Jo}, J.~{Huh}, J.~{Park}, B.~{Kim}, and J.~{Seo}.
\newblock Livegantt: Interactively visualizing a large manufacturing schedule.
\newblock {\em IEEE Transactions on Visualization and Computer Graphics},
  20(12):2329--2338, 2014. doi: {{%
10\hspace{.1pt}\discretionary{.}{%
}{.}\hspace{.4pt}1109\discretionary{/}{%
}{/}TVCG\hspace{.1pt}\discretionary{.}{%
}{.}\hspace{.4pt}2014\hspace{.1pt}\discretionary{.}{%
}{.}\hspace{.4pt}2346454}}


\bibitem{Lamport1978}
L.~Lamport.
\newblock Time, clocks, and the ordering of events in a distributed system.
\newblock {\em Communications of the ACM}, 21(7):558--565, July 1978. doi: {{%
10\hspace{.1pt}\discretionary{.}{%
}{.}\hspace{.4pt}1145\discretionary{/}{%
}{/}359545\hspace{.1pt}\discretionary{.}{%
}{.}\hspace{.4pt}359563}}


\bibitem{Leblanc1990}
T.~J. Leblanc, J.~M. Mellor-Crummey, and R.~J. Fowler.
\newblock {Analyzing parallel program executions using multiple views}.
\newblock {\em Journal of Parallel and Distributed Computing}, 9(2):203--217,
  jun 1990. doi: {{%
10\hspace{.1pt}\discretionary{.}{%
}{.}\hspace{.4pt}1016\discretionary{/}{%
}{/}0743\discretionary{%
}{-}{-}7315\discretionary{%
}{(}{(}90\discretionary{)}{%
}{)}90046\discretionary{%
}{-}{-}R}}


\bibitem{maguire2013visual}
E.~Maguire, P.~Rocca-Serra, S.-A. Sansone, J.~Davies, and M.~Chen.
\newblock Visual compression of workflow visualizations with automated
  detection of macro motifs.
\newblock {\em IEEE transactions on visualization and computer graphics},
  19(12):2576--2585, 2013.

\bibitem{lassen}
B.~McCandless.
\newblock Lassen.
\newblock codesign.llnl.gov/lassen.php, 2013.

\bibitem{McKenna2016}
S.~McKenna, D.~Staheli, C.~Fulcher, and M.~Meyer.
\newblock Bubblenet: A cyber security dashboard for visualizing patterns.
\newblock {\em Comput. Graph. Forum}, 35(3):281–290, June 2016.

\bibitem{Muelder2009}
C.~Muelder, F.~Gygi, and K.-L. Ma.
\newblock Visual analysis of inter-process communication for large-scale
  parallel computing.
\newblock {\em IEEE Transactions on Visualization and Computer Graphics,
  Proceedings of InfoVis}, 15(6):1129--1136, 2009. doi: {{%
10\hspace{.1pt}\discretionary{.}{%
}{.}\hspace{.4pt}1109\discretionary{/}{%
}{/}TVCG\hspace{.1pt}\discretionary{.}{%
}{.}\hspace{.4pt}2009\hspace{.1pt}\discretionary{.}{%
}{.}\hspace{.4pt}196}}


\bibitem{nagel1996vampir}
W.~E. Nagel, A.~Arnold, M.~Weber, H.-C. Hoppe, and K.~Solchenbach.
\newblock Vampir: Visualization and analysis of mpi resources.
\newblock {\em Supercomputer}, 1996.

\bibitem{Osmari2014SmartTraces}
D.~K. Osmari, H.~T. Vo, C.~T. Silva, J.~L. Comba, and L.~Lins.
\newblock Visualization and analysis of parallel dataflow execution with smart
  traces.
\newblock In {\em 27th SIBGRAPI Conference on Graphics, Patterns and Images
  (SIBGRAPI)}, pp. 165--172, Aug 2014. doi: {{%
10\hspace{.1pt}\discretionary{.}{%
}{.}\hspace{.4pt}1109\discretionary{/}{%
}{/}SIBGRAPI\hspace{.1pt}\discretionary{.}{%
}{.}\hspace{.4pt}2014\hspace{.1pt}\discretionary{.}{%
}{.}\hspace{.4pt}2}}


\bibitem{prolific}
Prolific.
\newblock Prolific.
\newblock https://www.prolific.co/, 2021.
\newblock (accessed 29 March 2021).

\bibitem{Reissmann2017GrainGraphs}
N.~Reissmann, M.~Jahre, and A.~Muddukrishna.
\newblock Towards aggregated grain graphs.
\newblock In {\em Fourth International Workshop on Visual Performance
  Analysis}, VPA 17, Nov. 2017.

\bibitem{flask}
A.~Ronacher.
\newblock Flask: web development, one drop at a time.
\newblock http://flask.pocoo.org/, 2021.
\newblock (accessed 29 March 2021).

\bibitem{amg}
J.~Ruge and K.~St\"{u}ben.
\newblock Algebraic multigrid ({AMG}).
\newblock In S.~McCormick, ed., {\em Multigrid Methods}, vol.~3 of {\em
  Frontiers in Applied Mathematics}. SIAM, 1987.

\bibitem{Schaubschlager2003DeWiz}
C.~Schaubschl\"{a}ger, D.~Kranzlm\"{u}ller, and J.~Volkert.
\newblock Event-based program analysis with {DeWiz}.
\newblock In {\em Proceedings of the Fifth International Workshop on Automated
  Debugging AADEBUG2003}, AADEBUG, 2003.

\bibitem{experiment_platform}
C.~Scully-Allison and K.~Isaacs.
\newblock Communication pattern experiment.
\newblock https://dry-beyond-38655.herokuapp.com/demographics, 2021.
\newblock (accessed 29 March 2021).

\bibitem{Sigovan2013}
C.~Sigovan, C.~W. Muelder, and K.-L. Ma.
\newblock Visualizing large-scale parallel communication traces using a
  particle animation technique.
\newblock {\em Computer Graphics Forum}, 32(3pt2):141--150, 2013. doi: {{%
10\hspace{.1pt}\discretionary{.}{%
}{.}\hspace{.4pt}1111\discretionary{/}{%
}{/}cgf\hspace{.1pt}\discretionary{.}{%
}{.}\hspace{.4pt}12101}}


\bibitem{pf3d}
C.~H. Still, R.~L. Berger, A.~B. Langdon, D.~E. Hinkel, L.~J. Suter, and E.~A.
  Williams.
\newblock Filamentation and forward brillouin scatter of entire smoothed and
  aberrated laser beams.
\newblock {\em Physics of Plasmas}, 7(5):2023--2032, 2000.

\bibitem{Stitz2016}
H.~Stitz, S.~Luger, M.~Streit, and N.~Gehlenborg.
\newblock Avocado: Visualization of workflow-derived data provenance for
  reproducible biomedical research.
\newblock {\em Computer Graphics Forum}, 35(3):481–490, June 2016.

\end{thebibliography}
\end{document}